\documentclass[twoside,fleqn]{article}
\usepackage{espcrc2}
\usepackage{graphicx}

\title{
\vspace{-1.0cm}
{\sf \small \rightline{IFIC/03-43, FTUV/03-0930}}
\bigskip
{\Large  Phenomenology of the $\langle\mathrm{VVP}\rangle$ Green's 
function within the Resonance Chiral Theory}
\thanks{Talk given by P.~D.~Ruiz-Femen\'\i a at the High-Energy Physics
International Conference in Quantum Chromodynamics 
(QCD 03), Montpellier, France, 2--8 July 2003.}}
\author{P. D. Ruiz-Femen\'\i a, A. Pich and J. Portol\'es
\address{
       {\em Instituto de F\'\i sica Corpuscular, Universitat de Val\`encia,} \\
       {\em Apartat Correus 22085, E-46071 Val\`encia, Spain}}
       }

\begin{document}

\newcommand{\nn}{\nonumber}
\newcommand{\mev}{\mbox{\rm MeV}}
\newcommand{\gev}{\mbox{\rm GeV}}
\newcommand{\eqn}[1]{(\ref{#1})}
\newcommand{\MSb}{{\overline{MS}}}
\newcommand{\ep}{\epsilon}
\newcommand{\IM}{\mbox{\rm Im}}
\newcommand{\lsim}{\stackrel{<}{_\sim}}
\newcommand{\gsim}{\stackrel{>}{_\sim}}

\newcommand{\br}{\mbox{Br}\,}
\newcommand{\tr}{\mbox{Tr}\,}
\newcommand{\tppp}{\tau^-\to\pi\pi\pi\nu_\tau}
\newcommand{\tpppc}{\tau^-\to\pi^+\pi^-\pi^-\nu_\tau}
\newcommand{\tpppn}{\tau^-\to\pi^-\pi^0\pi^0\nu_\tau}
\newcommand{\kbar}{\overline{K}^0}
\newcommand{\amp}{{\cal M}_{\kmm}}
\newcommand{\diag}{\mbox{diag}}
\newcommand{\rms}{\rm\scriptsize}
\newcommand{\rmf}{\rm\footnotesize}
\newcommand{\beq}{\begin{equation}}
\newcommand{\eeq}{\end{equation}}
\newcommand{\Frac}[2]{\frac{\displaystyle #1}{\displaystyle #2}}
\newcommand{\Oa}{{\cal O} (\alpha_{s}^3)}
\def\mapright#1#2{\smash{
     \mathop{-\!\!\!-\!\!\!-\!\!\!\longrightarrow}\limits^{#1}_{#2}}}

\begin{abstract}
We analyse the
odd-intrinsic-parity effective Lagrangian of QCD valid for processes
involving one pseudoscalar with two vector mesons described in terms of 
antisymmetric tensor fields.
Substantial information on the odd-intrinsic-parity couplings is obtained by
constructing the vector-vector-pseudoscalar Green's
three-point function, at leading order in $1/N_C$,
and demanding that its short-distance 
behaviour matches the corresponding OPE result. 
The QCD constraints thus enforced allow us to
predict the decay amplitude $\omega\to \pi \gamma$, the ${\cal O}(p^6)$
corrections to $\pi\to\gamma\gamma$
and the slope parameter in  
$\pi\to\gamma\gamma^*$. 
\end{abstract}

\maketitle

\newcommand{\jhep}[3]{{\it JHEP }{\bf #1} (#2) #3}
\newcommand{\nc}[3]{{\it Nuovo Cim. }{\bf #1} (#2) #3}
\newcommand{\npb}[3]{{\it Nucl. Phys. }{\bf B #1} (#2) #3}
\newcommand{\npps}[3]{{\it Nucl. Phys. }{\bf #1} {\it(Proc. Suppl.)} (#2) #3}
\newcommand{\plb}[3]{{\it Phys. Lett. }{\bf B #1} (#2) #3}
\newcommand{\pr}[3]{{\it Phys. Rev. }{\bf #1} (#2) #3}
\newcommand{\prd}[3]{{\it Phys. Rev. }{\bf D #1} (#2) #3}
\newcommand{\prl}[3]{{\it Phys. Rev. Lett. }{\bf #1} (#2) #3}
\newcommand{\prep}[3]{{\it Phys. Rep. }{\bf #1} (#2) #3}
\newcommand{\zpc}[3]{{\it Z. Physik }{\bf C #1} (#2) #3}
\newcommand{\sjnp}[3]{{\it Sov. J. Nucl. Phys. }{\bf #1} (#2) #3}
\newcommand{\jetp}[3]{{\it Sov. Phys. JETP }{\bf #1} (#2) #3}
\newcommand{\jetpl}[3]{{\it JETP Lett. }{\bf #1} (#2) #3}
\newcommand{\ijmpa}[3]{{\it Int. J. Mod. Phys. }{\bf A #1} (#2) #3}
\newcommand{\hepph}[1]{{\tt hep-ph/#1}} 
\newcommand{\hepth}[1]{{\tt hep-th/#1}} 
\newcommand{\heplat}[1]{{\tt hep-lat/#1}}



\section{Introduction}

Effective field theories of QCD have provided efficient ways to explore
hadron dynamics in those regimes where we are not able to solve the full theory.
In the very
low-energy domain, chiral perturbation theory ($\chi$PT)
\cite{WE79,GH84}
has achieved a remarkable success in describing the
strong interactions among pseudoscalar mesons. Moving up to the 1 GeV region
the effects of vector resonances become
dominant and must be accommodated in the theory. Several works 
\cite{EG89,EG89a}
have provided a sound procedure to include resonance states within 
the chiral 
framework, namely the Resonance Chiral Theory (R$\chi$T). 
As the couplings entering the effective Lagrangian 
are not fixed by the symmetry alone, one should rely on the phenomenology 
or, alternatively, construct theoretical tools that could provide a
meaningful way to compare the results of the effective theory with
those of QCD. The pioneering work of Ref.~\cite{EG89a} indicated that 
the analysis of Green's functions and form factors of QCD currents
yields valuable information on the 
resonance sector. 
\par
Recently, several authors have pushed forward this direction, either
by using a Lagrangian with explicit resonance degrees of freedom
or within the framework of the lowest meson dominance (LMD) approximation to
the large number of colours ($N_C$) limit of QCD 
\cite{KN01,BG03,M95,M97,KP99}. In
particular, the authors of Ref.~\cite{KN01} undertook a systematic study of 
several QCD three-point
functions which are free of perturbative contributions from QCD at 
short distances. 
Therefore, their OPE   
expansion should be
more reliable when descending to energies close to the resonance region.
Under
this hypothesis, it was shown \cite{KN01} that while the ansatz derived
from the LMD approach automatically incorporates the right short-distance
behaviour of QCD by construction, the same Green's functions as calculated
with a resonance Lagrangian, in the vector-field representation, are 
incompatible with the OPE outcome. Moreover the authors put forward that
these discrepancies cannot be repaired just by introducing 
local counterterms from the chiral Lagrangian ${\cal L}_{\chi}^{(6)}$, as 
it was done at ${\cal O}(p^4)$ \cite{EG89a}. This
result severely questions the usefulness of the resonance effective
theory beyond the initial work of Ref.~\cite{EG89a}, and deserves
further investigation. 
\par
With this aim, we have reanalysed 
the vector-vector-pseudoscalar
three-point function, this time with the vector mesons described
in terms of antisymmetric tensor fields. This
requires the introduction
of an odd-intrinsic-parity effective Lagrangian in the formulation of 
Ref.~\cite{EG89} containing
all allowed interactions between two vector objects (currents or resonances)
and one pseudoscalar meson. The details of the calculation can be
found in Ref.~\cite{RPP03}.

\section{R$\chi$T and the odd-intrinsic-parity sector}
\label{sec:RChPT}

The low-energy behaviour of QCD for the light quark sector
is ruled 
by the spontaneous breaking of chiral symmetry. The corresponding
effective realization of QCD describing the interaction between the
Goldstone fields is $\chi$PT 
given, at ${\cal O}(p^2)$, by 
\begin{equation}
{\cal L}_{\chi}^{(2)}=(F^2/4)\,\langle
u_{\mu}u^{\mu}+\chi^+\rangle\,.
\end{equation}
The inclusion of resonances as explicit degrees of freedom
in the chiral framework was carried out
in Ref.~\cite{EG89} 
for the even--intrinsic--parity sector (${\cal L}_{\mathrm{V}}$). 
For the odd--intrinsic--parity sector, three different 
sources might contribute to the 
$\langle\mathrm{VVP}\rangle$ Green's function~: 
\begin{itemize}
\item[(i)] 
the Wess-Zumino action $Z_{\mathrm{WZ}}[v,a]$ \cite{WZ71}, which is 
of ${\cal O}(p^4)$ and fulfills 
the chiral anomaly,
\item[(ii)]
chiral invariant $\epsilon_{\mu\nu\rho\sigma}$ terms 
involving vector mesons.
Within the antisymmetric formalism, the basis of 
odd-intrinsic-parity operators which comprise all possible 
vertices involving two vector resonances and one pseudoscalar (VVP), and 
vertices with one vector resonance and one external vector source plus one 
pseudoscalar (VJP) reads:
\begin{eqnarray}
{\cal O}_{\mbox{\tiny VJP}}^1 & = & \epsilon_{\mu\nu\rho\sigma}\,
\langle \, \{V^{\mu\nu},f_{+}^{\rho\alpha}\} \nabla_{\alpha}u^{\sigma}\,\rangle
\; \; , \nonumber\\[2mm]
{\cal O}_{\mbox{\tiny VJP}}^2 & = & \epsilon_{\mu\nu\rho\sigma}\,
\langle \, \{V^{\mu\alpha},f_{+}^{\rho\sigma}\} \nabla_{\alpha}u^{\nu}\,\rangle
\; \; , \nonumber\\[2mm]
{\cal O}_{\mbox{\tiny VJP}}^3 & = & i\,\epsilon_{\mu\nu\rho\sigma}\,
\langle \, \{V^{\mu\nu},f_{+}^{\rho\sigma}\}\, \chi_{-}\,\rangle
\; \; , \nonumber\\[2mm]
{\cal O}_{\mbox{\tiny VJP}}^4 & = & i\,\epsilon_{\mu\nu\rho\sigma}\,
\langle \, V^{\mu\nu}\,[\,f_{-}^{\rho\sigma}, \chi_{+}]\,\rangle
\; \; , \nonumber\\[2mm]
{\cal O}_{\mbox{\tiny VJP}}^5 & = & \epsilon_{\mu\nu\rho\sigma}\,
\langle \, \{\nabla_{\alpha}V^{\mu\nu},f_{+}^{\rho\alpha}\} u^{\sigma}\,\rangle
\; \; ,\nonumber\\[2mm]
{\cal O}_{\mbox{\tiny VJP}}^6 & = & \epsilon_{\mu\nu\rho\sigma}\,
\langle \, \{\nabla_{\alpha}V^{\mu\alpha},f_{+}^{\rho\sigma}\} u^{\nu}\,\rangle
\; \; , \nonumber\\[2mm]
{\cal O}_{\mbox{\tiny VJP}}^7 & = & \epsilon_{\mu\nu\rho\sigma}\,
\langle \, \{\nabla^{\sigma}V^{\mu\nu},f_{+}^{\rho\alpha}\} u_{\alpha}\,\rangle
\;\; ,
\label{eq:VJP}
\end{eqnarray}
\begin{eqnarray}
{\cal O}_{\mbox{\tiny VVP}}^1 & = & \epsilon_{\mu\nu\rho\sigma}\,
\langle \, \{V^{\mu\nu},V^{\rho\alpha}\} \nabla_{\alpha}u^{\sigma}\,\rangle
\; \; , \nonumber\\[2mm]
{\cal O}_{\mbox{\tiny VVP}}^2 & = & i\,\epsilon_{\mu\nu\rho\sigma}\,
\langle \, \{V^{\mu\nu},V^{\rho\sigma}\}\, \chi_{-}\,\rangle
\; \; , \nonumber\\[2mm]
{\cal O}_{\mbox{\tiny VVP}}^3 & = & \epsilon_{\mu\nu\rho\sigma}\,
\langle \, \{\nabla_{\alpha}V^{\mu\nu},V^{\rho\alpha}\} u^{\sigma}\,\rangle
\; \; , \nonumber\\[2mm]
{\cal O}_{\mbox{\tiny VVP}}^4 & = & \epsilon_{\mu\nu\rho\sigma}\,
\langle \, \{\nabla^{\sigma}V^{\mu\nu},V^{\rho\alpha}\} u_{\alpha}\,\rangle
\; \; . 
\label{eq:VVP}
\end{eqnarray}                       
The corresponding
resonance Lagrangian will 
thus be defined as
\begin{equation}
{\cal L}_V^{\mathrm{odd}} =
\sum_{a=1}^{7} \,\frac{c_a}{M_{V}}\,{\cal O}^a_{\mbox{\tiny{VJP}}} 
+
\sum_{a=1}^{4} \,d_a\,{\cal O}^a_{\mbox{\tiny{VVP}}}
\ .
\label{eq:Lano}
\end{equation}
\item[(iii)] 
the relevant operators in the ${\cal O}(p^6)$ Goldstone
chiral Lagrangian \cite{BGT02}. 
The successful 
resonance saturation of the chiral Lagrangian couplings at ${\cal O}(p^4)$
\cite{EG89} might translate naturally to ${\cal O}(p^6)$ couplings too,
implying that 
they are generated completely through
integration of vector resonances. Accordingly we do not include
${\cal L}_{\mathrm{odd}}^{(6)}$ in our evaluation. 
\end{itemize}
\par
In summary we will proceed in the following by considering the relevant 
effective resonance theory (ERT) given by~:
\begin{equation}
Z_{\mathrm{ERT}}[v,a,s,p]  = Z_{\mathrm{WZ}}[v,a] \, + \, 
Z_{\mathrm{V\chi}}^{\mathrm{odd}}[v,a,s,p] \,,
\label{eq:allZ}
\end{equation}
where $Z_{\mathrm{V\chi}}^{\mathrm{odd}}$ 
is generated by ${\cal L}_{\chi}^{(2)}$, ${\cal L}_{\mathrm{V}}$ 
and ${\cal L}_V^{\mathrm{odd}}$.

\section{Short-distance information on the odd-intrinsic-parity couplings}
\label{sec:short}

The
vector-vector-pseudoscalar QCD 
three-point function $\langle\mathrm{VVP}\rangle$ is built from
the octet vector current and the octet pseudoscalar density,
\begin{eqnarray}
(\Pi_{\mathrm{VVP}})^{(abc)}_{\mu\nu}&&\!\!\!\!\!\!\!\!\!\!\!(p,q)=\int  
d^4x\int d^4y
\,e^{i(p\cdot x+q\cdot y)}\nonumber\\[2mm]
\times&&\!\!\!\!\!\!\!\!\!\!\!
\langle 0|\,T\,[\,V^a_{\mu}(x)\,V^b_{\nu}(y)\,P^c(0)\,]\, |0\rangle 
\nonumber\\[2mm]
=&&\!\!\!\!\!\!\!\!\!\!\!\epsilon_{\mu\nu\alpha\beta}\,
p^{\alpha}q^{\beta}d^{abc}\,\Pi_{\mathrm{VVP}}(p^2,q^2,r^2)\,,
\label{eq:full3point}
\end{eqnarray}
with the four-vector $r=-(p+q)$. 

When both momenta $p,q$ in $\Pi_{\mathrm{VVP}}$ become simultaneously
large, the QCD calculation within the OPE framework gives, in the 
chiral limit and up to corrections
of ${\cal O}(\alpha_s)$, \cite{M95}:
\begin{eqnarray}
\lim_{\lambda \to \infty} \Pi_{\mathrm{VVP}}&&\!\!\!\!\!\!\!\!\!\!\!((\lambda p)^2,(\lambda q)^2,
(\lambda r)^2)
\nonumber\\[1mm]
=&&\!\!\!\!\!\!\!\!\!\!\!-\frac{\langle \bar{\psi}\psi \rangle_0 }{2\lambda^4}\,
\frac{p^2+q^2+r^2}{p^2q^2r^2}+{\cal O}\left(\frac{1}{\lambda^6}\right)\, ,
\label{eq:short1}
\end{eqnarray}
where $\langle \bar{\psi}\psi \rangle_0$ is the single flavour bilinear
quark condensate. 
\begin{figure}[tb]
\begin{center}
\hspace*{-0.4cm}
\includegraphics[angle=0,width=0.52\textwidth]{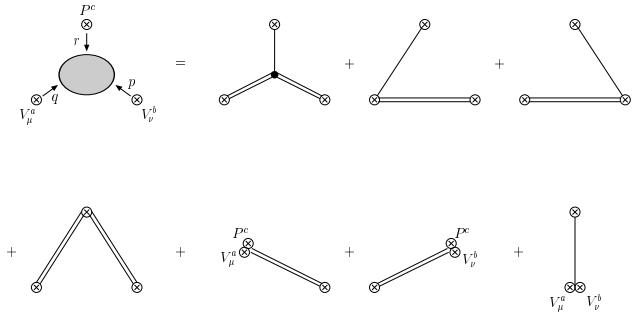}
\vspace*{-0.7cm}
\caption[]{\label{fig1} Diagrams entering the calculation of
the VVP 3-point function with the ERT action. Double lines
represent vector resonances, single lines are short for pseudoscalar
mesons.}
\vspace*{-0.8cm}
\end{center}
\end{figure}
At leading order in the $1/N_C$ expansion of QCD, the
three-point correlator in the effective resonance theory given by
$Z_{\mathrm{ERT}}$ is evaluated from the tree-level diagrams
shown in Fig.~\ref{fig1}. 
The LMD approximation, which assumes that 
a single resonance in each channel saturates the requirements of QCD,
is sufficient to satisfy the short--distance
constraint (\ref{eq:short1}) up to
order $1/\lambda^4$, provided the
following conditions among the ${\cal L}_V^{\mathrm{odd}}$ couplings hold:
\begin{eqnarray}
4 \, c_3+c_1&=&0 \; \; \; ,
\nonumber\\[1mm]
c_1-c_2+c_5 &=& 0 \; \; \; ,
\nonumber\\[1mm]
c_5-c_6&=& \frac{N_C}{64\pi^2}\frac{M_V}{\sqrt{2}F_V} \; \; \; ,
\nonumber\\[1mm]
d_1 + 8 \, d_2&=& -\frac{N_C}{64\pi^2}\frac{M_V^2}{F_V^2} \, + 
\, \frac{F^2}{4F_V^2}  \; \; \; ,
\nonumber\\[0mm]
d_3&=&-\frac{N_C}{64\pi^2}\frac{M_V^2}{F_V^2} \, + \, \frac{F^2}{8F_V^2}\,.
\label{eq:cond}
\end{eqnarray}
As 
the couplings of the Effective Lagrangian do not depend on the masses
of the Goldstone fields the constraints above
apply for non--zero pseudoscalar masses too.
\par
Actually our $\langle\mathrm{VVP}\rangle$ three-point function
fully reproduces the LMD ansatz suggested in Ref.~\cite{M95}~:
\begin{eqnarray}
\Pi_{\mathrm{VVP}}^{\mathrm{res}}=-\frac{\langle \bar{\psi}\psi \rangle_0 }{2}\, 
\frac{(p^2+q^2+r^2)-\frac{N_C}{4\pi^2}\frac{M_V^4}{F^2}}
{(p^2-M_V^2)(q^2-M_V^2)r^2}\, ,
\label{eq:VVPantsaz}
\end{eqnarray}
which has been successfully tested in previous works \cite{M95,KP99}. 
The authors of Ref.~\cite{KN01} found that the same agreement with 
the short-distance QCD behaviour could not be reached
working with the resonance Lagrangian 
in the vector representation, not even at the expense of introducing local
contributions from the ${\cal O}(p^6)$ chiral Lagrangian.
They then suggested that the problem may be 
inherent to the effective Lagrangian approach and unlikely to be
fixed just by using other representations for the resonance fields; 
our result, derived in the antisymmetric tensor-field 
formulation with an odd-intrinsic-parity sector, 
contradicts this assertion, at
least in what concerns the $\langle\mathrm{VVP}\rangle$ Green's function.

\section{Phenomenology of intrinsic-parity violating processes}
\label{sec:pheno}

\subsection{\mbox{\boldmath$\omega\to\pi\gamma$}}
\vspace*{0.2cm}

At tree-level, the intrinsic-parity violating transition $\omega\to\pi\gamma$ 
receives contributions from
both the VJP and VVP terms of ${\cal L}_V^{\mathrm{odd}}$ (direct and
$\rho$-mediated diagrams respectively). 
If we plug in the QCD constraints, Eq.~(\ref{eq:cond}),
we find a full prediction for 
this process~:
\begin{eqnarray}
\Gamma(\omega\to\pi\gamma)\!\!\!\!\!\!&&\!\!\!\!=\frac{\alpha}{192}\,M_{\omega}
\bigg( 1-\frac{m^2_{\pi}}{M_{\omega}^2}\bigg)^3\,
\nonumber\\[2mm]
\times\!\!\!\!&&\!\!\!\!\!\!\bigg[\,\frac{N_C}{4\pi^2}\,\frac{M_{\omega}^2}{F^2}
\,-\frac{M_{\omega}^2}{M_{V}^2} \left( 1 + 
\Frac{m_{\pi}^2}{M_\omega^2} \, \right)
\,\bigg]^2.
\label{eq:Gamma_w}
\end{eqnarray}  
The direct and the $\rho$ exchange
diagrams almost contribute to similar extent 
to this process. This means that contrary to what we would expect from 
VMD, the $\omega\rho\pi$ coupling does not saturate  
the decay $\omega\to\pi\gamma$. This has immediate consequences to
other channels where VMD alone was thought to be the
relevant mechanism of decay, as in $\omega\to\pi^+\pi^-\pi^0$, 
where the direct
amplitude competes in size with the intermediate meson exchange
term \cite{RPP03}.  

Varying the parameter $F$ from the bare value
$F_0\simeq 87$ MeV to the dressed one (i.e. the pion decay constant),
$F_{\pi} \simeq 92.4$ MeV \cite{PDG}, we get
that $\Gamma(\omega\to\pi\gamma)$ ranges from 0.703~MeV to 0.524~MeV,
with $M_V=M_{\rho}=771.1$ MeV and
$M_{\omega}=782.6$ MeV \cite{PDG}. 
This 5--30\% deviation from
the experimental value, 
$\Gamma(\omega\to\pi\gamma)|_{\mathrm{exp}}=(0.734\pm 0.035)\,\mathrm{MeV}$,  
is in accordance with the expected size of next-to-leading $1/N_C$ corrections. 
Also the $\rho\to\pi\gamma$ decay widths, related with $\omega\to\pi\gamma$
by a SU(3)$_{\mathrm{V}}$-symmetry factor, are extracted from the analysis
\cite{RPP03}.

\subsection{\mbox{\boldmath $\pi\to\gamma\gamma$}}
\vspace*{0.2cm}

In the chiral limit, the amplitude for the $\pi\to\gamma\gamma$ process
is non-vanishing and exactly predicted by
the ABJ anomaly. 
The odd-intrinsic-parity interactions among vector resonances
introduced in Section \ref{sec:RChPT} 
generate ${\cal O}(p^6)$ chiral corrections to this
process. 
Only the two-resonance driven diagram survives after the short-distance
conditions are applied. 
The correction induced into the $\pi \to \gamma \gamma$ width gives~:
\begin{equation}
\Gamma(\pi\to\gamma\gamma)=\frac{\alpha^2}{64 \, \pi^3 \, F^2}m_{\pi}^3
\left[ \, 1 \, - \, \Delta \, \right]^2 \; \; ,
\end{equation}
where
\begin{equation}
\Delta \; = \; 
\Frac{4\pi^2}{3} \, \Frac{F^2}{M_V^2} \; 
\Frac{m_{\pi}^2}{M_{V}^2} \,\, \simeq \; \; 0.006 \; \; \; .
\end{equation}
This result provides a tiny 1$\%
$ correction to the width, and it is
perfectly compatible with the experimental uncertainty, 
$\Gamma(\pi\to\gamma\gamma)|_{\mathrm{exp}}=(7.7\pm 0.6)$ eV.

\subsection{\mbox{\boldmath $\pi\to\gamma\gamma^*$}}
\vspace*{0.2cm}

The $\pi\to\gamma\gamma^*$ amplitude 
is usually written as a slope parameter $\alpha$ which modifies the on-shell
behaviour:
\begin{equation}
{\cal M}_{\pi\to\gamma\gamma^*}\,=\,
{\cal M}_{\pi\to\gamma\gamma}
\left(1+\alpha\,k^{*2}\right)\,,
\label{eq:alphaodd}
\end{equation}
where $k^*$ is the off-shell photon momentum. The interactions contained 
in ${\cal L}_V^{\mathrm{odd}}$ yield a contribution to the parameter
$\alpha$ that amounts
$$
\alpha^{\mathrm{odd}}\,=\,\frac{1}{M_V^2}
\left[1-\frac{4\pi^2F^2}{N_CM_V^2}\,\right]\,\simeq \,1.36\,\,\mathrm{GeV}^{-2} \,,
$$
which is smaller than the VMD estimate, $\alpha^{\mathrm{VMD}}=1/M_V^2\simeq
1.68$ GeV$^{-2}$. The chiral loops contributions to this slope,
$\alpha^{\chi}\simeq \,0.26\,\,\mathrm{GeV}^{-2}$
were calculated
in Ref.~\cite{BB90}. We can add both contributions
to get $m_{\pi}^2\alpha\simeq 0.029$, to
be compared with the averaged value $m_{\pi}^2\alpha|_{\mathrm{exp}}=
0.032\pm 0.004$ in the PDG~\cite{PDG}. 
$\alpha^{\mathrm{odd}}$ has been extended beyond the LMD approximation by 
the inclusion of a second vector resonance into the 
$\langle\mathrm{VVP}\rangle$ ansatz,
Eq.~(\ref{eq:VVPantsaz}), in Ref.~\cite{KN01}. The latter is in fact
needed to have the right $1/k^{*2}$ behaviour for large
$k^*$ \cite{ximo,LB79} in the form factor
${\cal F}_{\pi\gamma\gamma^*}(k^*)$.




\vspace*{0.3cm}
\noindent
{\bf Acknowledgements}

We wish to thank S. Narison and his team 
for the organization of the QCD 03 conference.
This work has been supported in part by TMR EURIDICE, EC Contract No. 
HPRN-CT-2002-00311, by MCYT (Spain) under grant FPA2001-3031, and
by ERDF funds from the EU.


\vspace*{-0.cm}
\begin{thebibliography}{99}

\bibitem{WE79} S.~Weinberg, Physica {\bf 96A} (1979) 327.

\bibitem{GH84} J.~Gasser and H.~Leutwyler, Ann. of Phys. (NY) {\bf 158}
               (1984) 142.


\bibitem{EG89} G.~Ecker, J.~Gasser, A.~Pich and E.~de~Rafael,
               Nucl. Phys. {\bf B321} (1989) 311.

\bibitem{EG89a} G.~Ecker, J.~ Gasser, H. Leutwyler, A. Pich
                and E. de Rafael, Phys. Lett. {\bf B223} (1989) 425.

\bibitem{KN01} M.~Knecht and A.~Nyffeler, Eur.\ Phys.\ J.\ C {\bf 21} (2001) 659.	

\bibitem{BG03} J.~Bijnens, E.~Gamiz, E.~Lipartia and J.~Prades,
	       JHEP {\bf 0304} (2003) 055.


\bibitem{M95} B.~Moussallam, Phys.\ Rev.\ D {\bf 51} (1995) 4939.

\bibitem{M97} B.~Moussallam, Nucl.\ Phys.\ B {\bf 504} (1997) 381.

\bibitem{KP99} M.~Knecht, S.~Peris, M.~Perrottet and E.~de Rafael,
Phys.\ Rev.\ Lett.\  {\bf 83} (1999) 5230.

\bibitem{RPP03} P.~D.~Ruiz-Femenia, A.~Pich and J.~Portol\'es,
JHEP {\bf 0307} (2003) 003.


\bibitem{WZ71} J.~Wess and B.~Zumino, Phys.\ Lett. {\bf B37} (1971) 95; \\
               E.~Witten, Nucl.\ Phys. B {\bf 223} (1983) 22.


\bibitem{BGT02} J.~Bijnens, L.~Girlanda and P.~Talavera,
		     Eur.\ Phys.\ J.\ C {\bf 23} (2002) 539.


\bibitem{PDG} K.~Hagiwara {\it et al.}  
[Particle Data Group Collaboration],
Phys.\ Rev.\ D {\bf 66} (2002) 010001.


\bibitem{BB90} J.~Bijnens, A.~Bramon and F.~Cornet,
Phys.\ Rev.\ Lett.\  {\bf 61} (1988) 1453.

\bibitem{ximo} Joaquim Prades, private communication.

\bibitem{LB79} G.~P.~Lepage and S.~J.~Brodsky, 
Phys. Lett. {\bf B87} (1979) 359.

\end{thebibliography}
\end{document}